\documentclass[prl, aps, bibliography, twocolumn, reprint, showpacs, footnoteinbib]{revtex4-1}
\usepackage{graphicx}
\usepackage{amssymb}   
\usepackage{amsfonts}
\usepackage{amsmath}
\usepackage{dsfont}
\usepackage{dcolumn}   
\usepackage{bm}        
\usepackage[mathscr]{eucal}
\usepackage[dvipsnames]{xcolor}
\usepackage[colorlinks, linkcolor=OrangeRed,citecolor=RoyalBlue,urlcolor=NavyBlue]{hyperref}
\usepackage[all]{hypcap} 
\usepackage{mathtools}
\usepackage{wasysym}

\newcommand{\mc}{\mathcal}

\newcommand{\eq}[1]{\begin{align}#1\end{align}}

\begin{document}

\title{Quantum Mutual Information as a Probe for Many-Body Localization }
\author{Giuseppe De Tomasi}
\author{Soumya Bera}
\author{Jens H. Bardarson}
\author{Frank Pollmann}

\affiliation{Max-Planck-Institut f\"ur Physik komplexer Systeme, N\"othnitzer Stra{\ss}e 38,  01187-Dresden, Germany}

\begin{abstract}
We demonstrate that the quantum mutual information (QMI) is a useful probe to study many-body localization (MBL). First, we focus on the detection of a metal--insulator transition for two
 different models, the noninteracting Aubry-Andr\'e-Harper model and the spinless fermionic disordered Hubbard chain. We find that the QMI in the localized phase decays exponentially with the distance between the regions 
traced out, allowing us to define a correlation length, which converges to the localization length in the case of one particle. 
Second, we show how the QMI can be used as a dynamical indicator to distinguish an Anderson insulator phase from an MBL phase. 
By studying the spread of the QMI after a global quench from a random product state, we show that the QMI does not spread in the Anderson insulator phase but grows logarithmically in time in the MBL phase.
\end{abstract}

\maketitle

{\it Introduction}---In the early sixties, Mott and Twose ~\cite{Mott61}, following Anderson's work  
~\cite{Anderson58},  conjectured that in one dimensional systems 
all single particle eigenstates are localized for any amount of uncorrelated disorder. This statement was given a mathematically rigorous proof by Gol'dshtein et al. \cite{Gol77} in the seventies.
Since the localization of all single particles eigenstates implies no transport, the resulting phase is a perfect insulator---the Anderson insulator ~\cite{Lan70, Ishii73}.
Afterwards, the problem of including interaction was studied extensively ~\cite{Alt82,Flei80, Gor05}, culminating in the seminal work of  Basko, Aleiner and Altshuler ~\cite{Basko06} 
demonstrating the possible existence of a metal-insulator transition at finite temperature in the presence of both disorder and interaction. 
This result has brought new emphasis and stimulated extensive research on the resulting many-body localization (MBL). 
The presence of a metal-insulator transition has been confirmed in several works ~\cite{Pr08, Pal10,  
Ros11, De13, Lev14, FP14, Be15, luitz15, Rajeev15, Rajeev16}, which also underline the ergodicity breaking in the MBL 
phase.
Moreover, new advancements of controlled experimental techniques allowed the first evidence of the existence of a localized 
phase and the presence of a transition ~\cite{Schreiber15, Smith2016, Choi16, Bordia16}. 
Nevertheless, one of the issues in the experiments has been to distinguish an Anderson insulator phase from an MBL phase. The growth of the entanglement entropy after a global quench shows different 
behavior between the two phases. In the Anderson insulator phase it saturates and in the MBL phase it grows 
logarithmically ~\cite{Ba12,Aba13,Pr08}; however, measuring entanglement entropy in an experimental setup is challenging 
due to its nonlocal nature ~\cite{Islam2015}.

In this work we propose the quantum mutual information (QMI) between two small spatially separated regions as a possible quantity that can in principle be used in an experimental setup to detect the transition and to distinguish between an Anderson insulator and an MBL phase,
without the need to compute an extensive many--body density matrix ~\cite{Amico08}.
\begin{figure}[tb!]
\includegraphics[width=0.95\columnwidth]{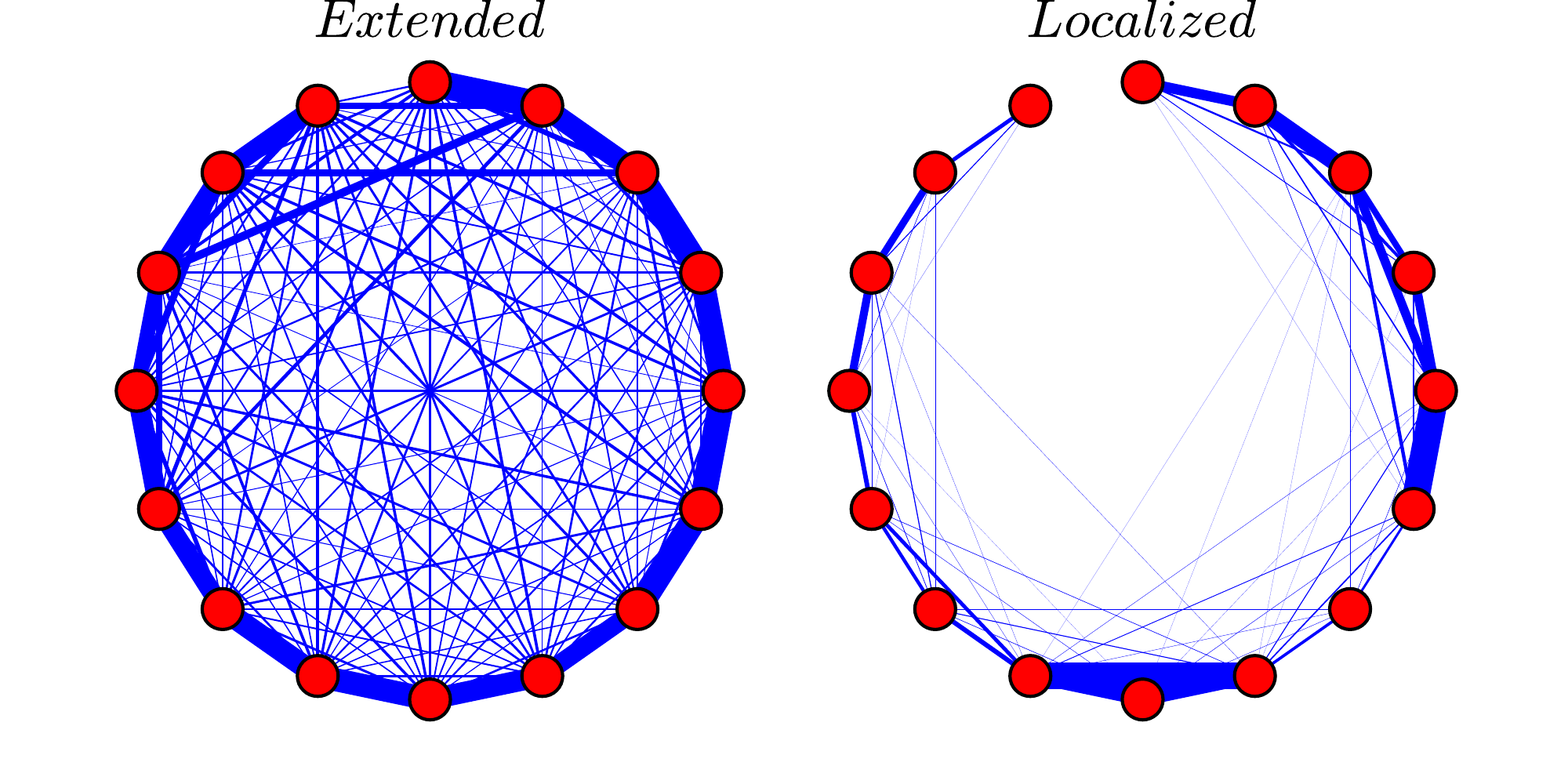}
\caption{Qualitative behavior of the QMI in the two different phases of the interacting disorder model $\mathcal{H}$ \eqref{eq:Ham} for a fixed disorder configuration. $W=1$ (left) and $W=5$ (right). The red dots represent the sites of the chain and the thickness of the blue bonds between sites $\{i,j\}$ is proportional 
to the magnitude of $\frac{\mathcal{I}([i],[j])}{\max_{i,j} \mathcal{I}([i],[j])}$ averaged over 16 eigenstates in the middle of the spectrum.}
\label{fig:Fancy}
\end{figure}
Several quantities have been borrowed from quantum information theory to characterize the extended and the localized phase as well as to detect the transition \cite{luitz15,Be16, Pr08, Ie15, Bela13,Gera16, Enss14, Dev15, Berk12}.  
The use of quantum information theory tools (i.e., entanglement entropy, R\'enyi entropy, concurrence, quantum mutual information) has been shown to be a resounding resource to study quantum critical points and different phases in
strongly correlated systems ~\cite{Amico08}. 
The mutual information measures the total amount of classical and quantum correlations in the system and has been successfully used to study phase transitions ~\cite{Mat96,Mat00,Alc13,Moore14, 
Melko10,Vid12, Chen10, Anf05, Ste14, Has11, Jae12, Cirac08}. We study the QMI between two sites in two different models.
The first one is the Aubry-Andr\'e-Harper (AAH) model ~\cite{Au80}, which is a one dimensional model (1D) of noninteracting fermions subject to a quasi-periodic potential, known to have a metal-insulator transition. 
The second is the paradigmatic model of 1D interacting spinless fermions subject to an uncorrelated random potential, which is believed to have an MBL transition ~\cite{luitz15, Pal10,Be15}.
The computation of the QMI between two sites involves only two point correlation functions and can thus in principle be 
measured in experiments ~\cite{Walborn2006, Jurcevic2014, Gross15,Islam2015}.

First we show that QMI in highly excited eigenstates decays exponentially with the distance between two sites in the localized phase, but algebraically in the extended phase. Using the QMI we define a correlation length which diverges at the transition and in the limit of one particle 
can be related to the localization length. Second, studying the dynamics after a global quench, we show how the QMI spreads differently in an Anderson insulator, an MBL phase and an extended phase.

{\it Model}---We study the Hamiltonian
\begin{equation} 
\begin{split}
 \mathcal{H}  := & -\frac{t}{2}\sum_{j=1}^{L-1} c^\dagger_{j} c_{j+1} + h.c.  + \sum_{j=1}^{L}  h_j  \left ( n_j - \frac{1}{2}\right)\\
 &  + V\sum_{j=1}^{L-1}   \left ( n_j - \frac{1}{2}\right)   \left ( n_{j+1} - \frac{1}{2}\right) 
\end{split}
\label{eq:Ham}
\end{equation}
where $c_j^\dagger~(c_j)$ is the fermionic creation (annihilation) operator at
site $j$ and $n_j = c_j^\dagger c_j $, $\{h_j \}$  are random fields,  $t$ and $V$ are respectively the hopping and the interaction strength, 
$L$ the system size and $N = \frac{L}{2}$ the number of fermions. We consider two different cases that have 
a metal-insulator transition. First, the noninteracting AAH model, which is obtained from  $\mathcal{H}$ \eqref{eq:Ham} with $V=0$, $t=2$ and  $h_j=W \cos ( 2\pi j \phi + \alpha)$ where $\phi=\frac{1+\sqrt{5}}{2}$ 
is the golden ratio and $\alpha$ is a random phase uniformly distributed in $[0, 2\pi]$. The AAH model is known to have a metal-insulator transition at $W_c=2$ (extended phase for $W\le W_c$ and localized phase for $W>W_c$). The localization length close to the 
transition diverges as $\xi_{\text{loc}} \sim \log^{-1}\frac{W}{2} $ ~\cite{Au80}. 
Second, the spinless fermionic disordered Hubbard chain is obtained by choosing $t=V=1$, and $\{h_j\}$ independent random variables uniformly distributed in $[-W,W]$. 
This Hubbard chain is believed to have an MBL transition at a critical disorder strength $W_c = 3.5\pm 1$ \cite{luitz15, 
Pal10, Be15} (extended for $W < W_c$ and localized for $W>W_c$).

{\it Quantum Mutual Information}---The quantum mutual information for two spatial subsets of the system $\mc{A}, 
\mathcal{B}  \subseteq [1,L]$ is defined as ~\cite{Amico08}:
\eq{ 
\mathcal{I(A,B)} := S( \mathcal{A}) + S( \mathcal{B}) - S( \mathcal{A} \cup \mathcal{B})
}
where $S( \mathcal{A})$ is the Von Neumann entropy $S( \mathcal{A}) = - \text{Tr} [ \rho_A \log \rho_A ] $ with $\rho_A$ the reduced density matrix of the subset $\mathcal{A}$
calculated using an eigenstate of $\mathcal{H}$.  
Figure \ref{fig:Fancy} shows the typical behavior of $\mathcal{I}([i],[j])$ for a given disorder 
configuration in two different phases (extended/localized) for all possible combination of bonds $\{i,j\}$. The thickness of the lines that connect $i \leftrightarrow j$ represents the magnitude of 
$\frac{\mathcal{I}([i],[j])}{\max_{i,j} \mathcal{I}([i],[j])}$. In the extended phase (Fig.~\ref{fig:Fancy}, left panel) the strongest bonds are the first neighbors $\{i,i+1\}$ but nevertheless all the other combinations of bonds have almost the same magnitude 
indicating that in the extended phase all sites are entangled with each other. In contrast, in the localized phase (Fig.~\ref{fig:Fancy}, right panel) each site is mainly entangled with neighboring sites and the QMI is almost zero for distant sites. 

To quantify this behavior, we focus our study on $\mathcal{I}_j =  \mathcal{I}([1],[j])$, from which we can define a correlation length 
\eq{
\xi^{-1} := -\lim_{j\rightarrow \infty} \frac{1}{j}\overline{ \log \frac{\mathcal{I}_j}{\mathcal{I}_1} } = \lim_{j\rightarrow \infty} \xi_j^{-1},
\label{eq:xi} }
where the overline stands for disorder average.  We expect that in the localized phase $\mathcal{I}_j$ decays exponentially in $j$ ($\mathcal{I}_j \sim e^{-\frac{j}{\xi}}$), thus $\xi^{-1}$ will be
nonzero. Instead, in the extended phase we expect a decay of $\mathcal{I}_j$ slower than exponential, implying $\xi^{-1}$ is zero in the thermodynamic limit.
The exponential decay of $\mathcal{I}_j$ implies, via the Pinsker's inequality, that all two point correlation functions also decay exponentially with the distance ~\cite{Ie15}.
This definition of a correlation length is related to the single particle localization length $\xi_{\text{loc}}$, which is defined as ~\cite{Kramer93}
\eq{
\xi_{\text{loc}}^{-1} := -\lim_{j\rightarrow \infty} \frac{1}{j}\log \frac{ |\psi_j|}{|\psi_0|}
\label{eq:xi_loc}
}
with $\psi_j$ the single particle wave function evaluated at site $j$. Computing $\mathcal{I}_j$ for a system composed of one fermion ($N=1$) and assuming $\psi_j$ is a decaying function of $j$
$$\log \mathcal{I}_j \sim \log|\psi_j|^2 + \log \left ( 1 - \log |\psi_j|^2 + \log \frac{|\psi_0|^2}{1-|\psi_0|^2} \right )$$ 
for large $j$, implies
\eq{ 
\xi \sim 2 \xi_{\text{loc}}.
} 
As a further measure of the spread of the QMI we interpret $\left \{ p_j = \frac{\mathcal{I}_j}{ \sum_m \mathcal{I}_m } \right \}$ as the values of a discrete probability distribution 
and take its variance
\eq{
\sigma^2 := \sum_j j^2 p_j - \left ( \sum_j\ j p_j  \right )^2 .
\label{eq:sigma}
}
Since we expect $\mathcal{I}_j$ to decay exponentially fast with $j$ in the localized phase, $\overline{\sigma}$ should saturate with system size in this phase. However, it is important to note that
$\overline{\sigma}$ can still saturate for algebraically decaying $\mathcal{I}_j$ (i.e., $\mathcal{I}_j \sim \frac{1}{j^{3+\eta}}$ for any $\eta > 0$), thus this
quantity can only be used to detect a lower bound of the transition point.

\begin{figure}[tb]
\includegraphics[width=1.\columnwidth]{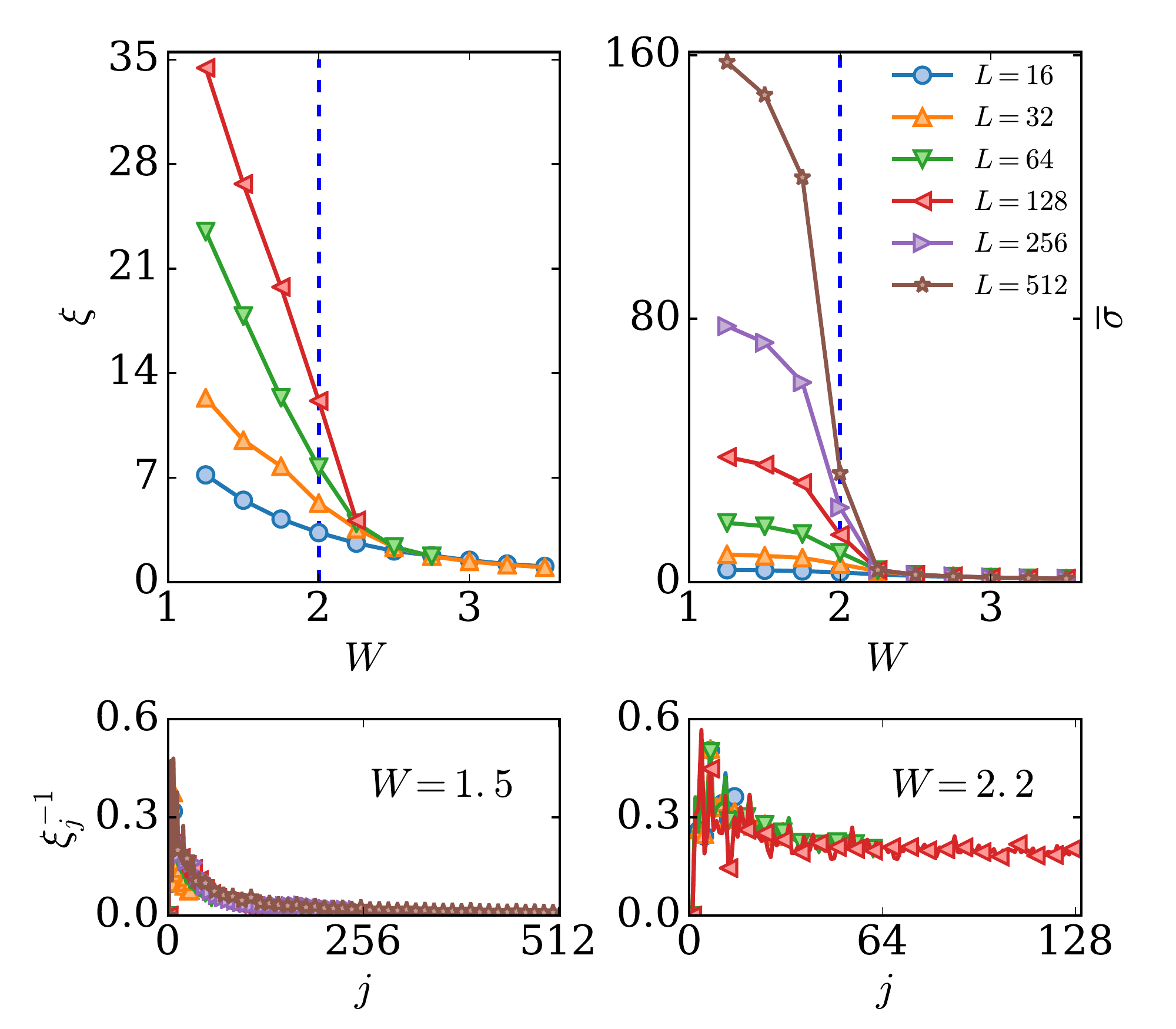}
\caption{(AAH-model) The upper left panel shows the localization length $\xi$ for different system sizes as a function of disorder strength $W$. The dashed line at $W_c =2$ represent the known transition point between extended and localized states ~\cite{Au80}. For values below $W_c$, $\xi$ increases
with system size while for values above $W_c$ it saturates. The right upper panel shows $\overline{\sigma}$ for different system sizes as a function of disorder strength $W$; for values of $W$ below $W_c$ $\overline{\sigma}$ grows with system size while for values above $W_c$ it saturates. 
The lower panels shows $\xi_j^{-1}$ in two different phases: for $W =1.5$ in the extended phase $\xi_j^{-1}$ goes to zero as a function of $j$, while for $W =2.2$ in the localized phase it saturates to a positive value implying a finite correlation length $\xi$.}
\label{fig:A_total}
\end{figure}
{\it Aubry-Andr\'e-Harper model}---We start by benchmarking our assumption on the behavior of the QMI in different 
phases for the AAH model.
We compute $\mathcal{I}_j$ for this model using a free fermion technique ~\cite{Pes09} for eigenstates of $\mathcal{H}$ constructed as a Slater determinant taking random single particle eigenstates, which implies an 
effective infinite temperature ensemble. 
The two lower panels of Fig.~\ref{fig:A_total} show $\xi_j^{-1}$ as a function of $j$, for two different values of $W$, in the extended phase
($W=1.5$) and in the localized phase ($W=2.2$). In the extended phase it decays to zero with a saturation point which scales as the inverse of the system size with a logarithmic
correction due to the normalization of the single particle wave functions $(\xi^{-1} \sim \frac{\log{L}}{L})$. 
In the localized phase,  $\xi^{-1}$ saturates to a nonzero value, leading to a finite correlation length.
The left upper panel of Fig.~\ref{fig:A_total} shows $\xi$  for different system sizes and different disorder strengths. In the localized phase for a fixed system size $L$,  $\xi$ was extrapolated from $\xi_j$ by averaging over the values of $j$ where it saturates, and in the extended phase we take $\xi=\xi_{j=L}$. As expected, in the extended phase $\xi$ increases with system size, 
while in the localized phase it saturates to a constant.  
The right upper panel of Fig.~\ref{fig:A_total} shows $\sigma $ averaged over disorder realizations for different disorder strengths and different system sizes. For values of $W$ greater than $W_c$, $\overline{\sigma}$ converges 
to a finite value, which implies that all the eigenstates are localized and have reached their maximum extension. However, for values below $W_c$, $\overline{\sigma}$  scales linearly with system size ($\overline{\sigma}\sim L$), with the 
consequence that $p_j \sim L^{-1}$, indicating that correlations are spread uniformly at any distance.
\begin{figure}[tb]
\includegraphics[width=1.0\columnwidth]{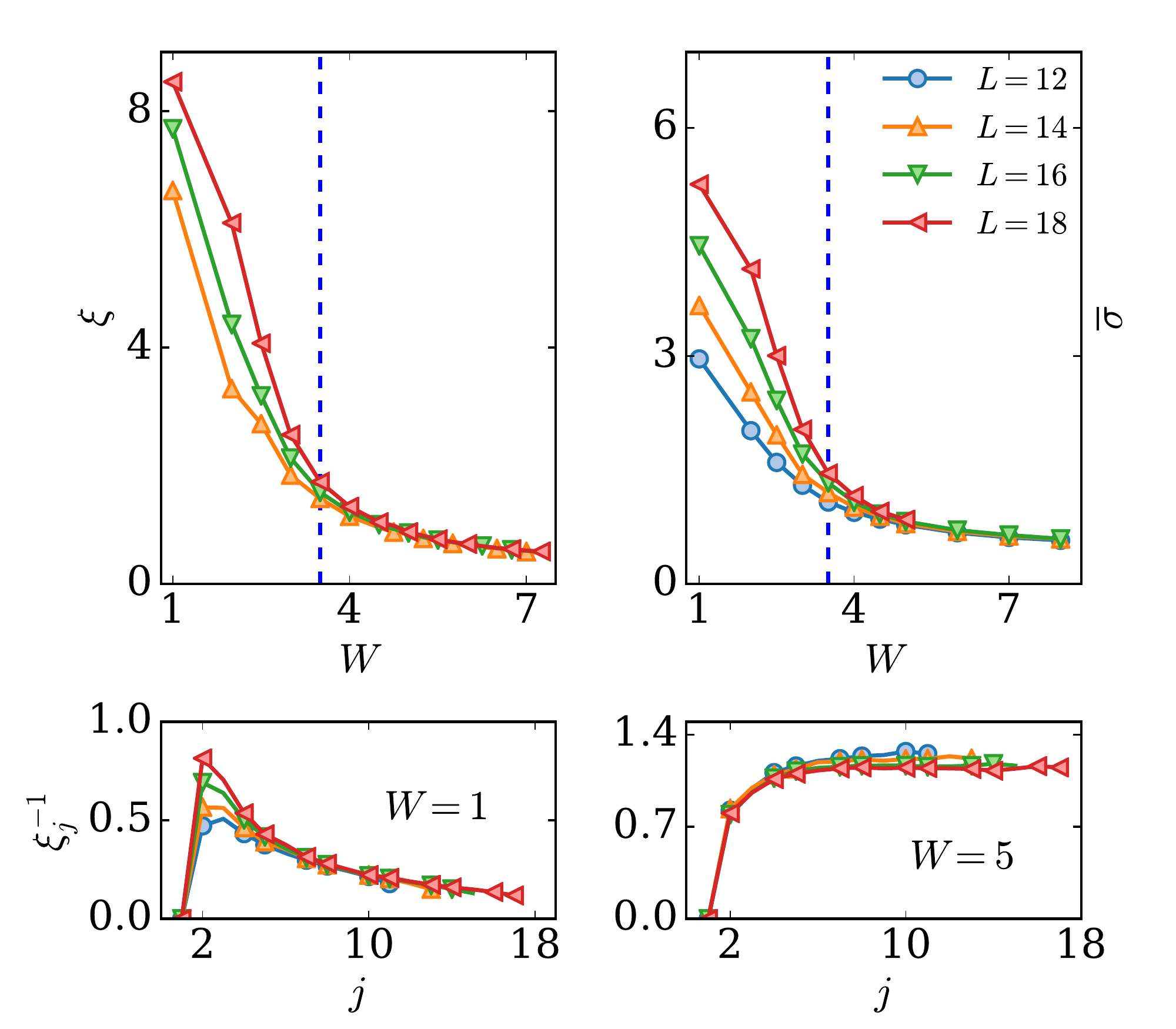}
\caption{(Spinless Hubbard chain) The top left panel shows the localization length $\xi$ for different system sizes as a function of disorder strength $W$.
The top right panel shows $\overline{\sigma}$ for different system sizes as a function of disorder strength $W$, for values $W< 4$ it grows with system size while for larger values it saturates. 
The vertical dashed line at $W_c = 3.5$ is the value for the expected transition \cite{luitz15, Pal10}.
The bottom lower panels show $\xi_j^{-1}$ in the two different phases. For $W=1$ in the extended phase, $\xi_j^{-1}$ goes to zero as a function of $j$, 
for $W>4$ in the localized phase it starts to saturate to a positive value implying a finite correlation length.}
\label{fig:MBL_Total}
\end{figure}

{\it Spinless Hubbard chain}---Having shown that the QMI captures the salient features of the metal--insulator 
transition in the AAH model, we now study $\mathcal{I}_j$ for the interacting problem that has an MBL transition.
For this model, we compute $\mathcal{I}_j$ using exact diagonalization for eigenstates in the middle of the spectrum. 
The lower panels of Fig.~\ref{fig:MBL_Total} show $\xi_j^{-1}$ for two different values of $W$. In the expected extended phase $(W=1)$, it goes to zero with increasing $j$ and in the MBL  phase $(W=5)$ it becomes constant for large $j$, indicating that the QMI
decays exponentially with $j$. As for the AAH-model, for values of $W$ where $\xi_j$ becomes a constant with respect to $j$ we average over those sites, and for values  of $W$ where $\xi_j$ decays uniformly with $j$ we take $\xi = \xi_{j=L}$. The left panel of Fig.~\ref{fig:MBL_Total} shows the extrapolation of the correlation length for different values of $W$ and for different $L$.
We note that for values $W < 4.0$, $\xi$ does not converge for available system sizes, but it increases with $L$ giving an indication of an extended phase and thus of a transition. 
As expected, $\xi$ is a monotonically decreasing function of $W$, implying stronger localization for larger disorder.
We also detect the extended and localized phases by studying $\overline{\sigma}$, as shown in the right upper panel of Fig.~\ref{fig:MBL_Total}. Its behavior is similar to the case of the AAH model. For values $W\le 4$, $\overline{\sigma}$ grows with 
$L$ ($\overline{\sigma} \sim L$), implying $p_j \sim L^{-1}$, so there is equal probability of finding correlation at 
any distance.  For $W > 4.0$, $\overline{\sigma}$ saturates with $L$ indicating the presence of the two different 
phases.

\begin{figure}[tb]
\includegraphics[width=1.0\columnwidth]{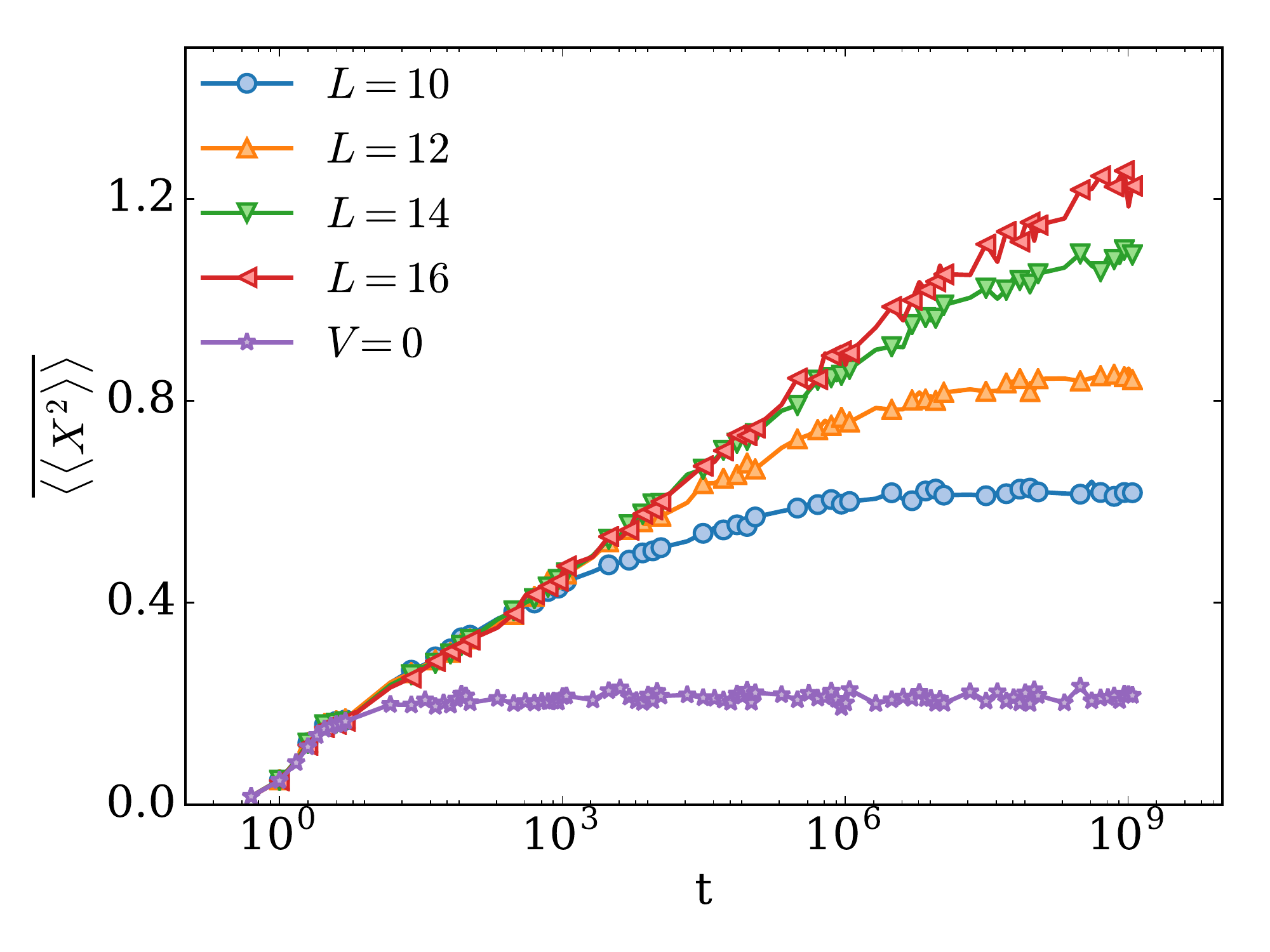}
\caption{$\overline{ \langle \langle X^2  \rangle \rangle}$ for different system sizes for $W=6$, and for $V = 0$ (non-interacting). For $V=0$ $\overline{\left \langle \langle X^2 \right \rangle \rangle}$ 
saturates at time of the order one and with system size. For $V \ne 0$, $\overline{\langle \langle X^2  \rangle \rangle} \sim \log (t)$.}
\label{fig:Time}
\end{figure}
{\it Unbounded spread of QMI}---We now show how  $\mathcal{I}_j$ can be used to distinguish between an Anderson 
insulator phase and an MBL phase. We perform a global
quench from a random product state $\big (\prod_{s=1}^N c^\dagger_{i_s} |0\rangle \big )$ and compute $\mathcal{I}_j$ as a function of time. We study the following quantity,
\eq{ 
\langle \langle X^2 \rangle \rangle := \sum_j j^2\mathcal{I}_j(t)  - \left ( \sum_j j  \mathcal{I}_j(t) \right )^2 .
}
This quantity allows us to detect the spread of information under time evolution.
At $t=0$ the initial product state has no entanglement and $\langle \langle X^2 \rangle \rangle$ is zero. With the increase of time its value increases.
Figure \ref{fig:Time} shows $\langle \langle X^2 \rangle \rangle$ as a function of time $t$ averaged over disorder and over random product states in the regime of strong localization $W=6$.
For $V = 0$ (Anderson model) it saturates at a time of the order one ($ \sim ( \text{hopping strength})^{-1}$) as one would expect in an Anderson insulator phase. In the MBL phase $(V \ne 0$) in contrast, it grows logarithmically, $\langle \langle X^2 \rangle \rangle \sim \log (t)$.
The logarithmic growth can be understood from the mechanism of dephasing induced by interaction \cite{Aba13} , in which the time needed to entangle separated portion of the system grows exponentially with their distance.
We tested this by calculating the minimum time such that $\overline{\mathcal{I}_j}(t)$ starts to be bigger than some fixed threshold,
\begin{figure}[tb]
\includegraphics[width=1.0\columnwidth]{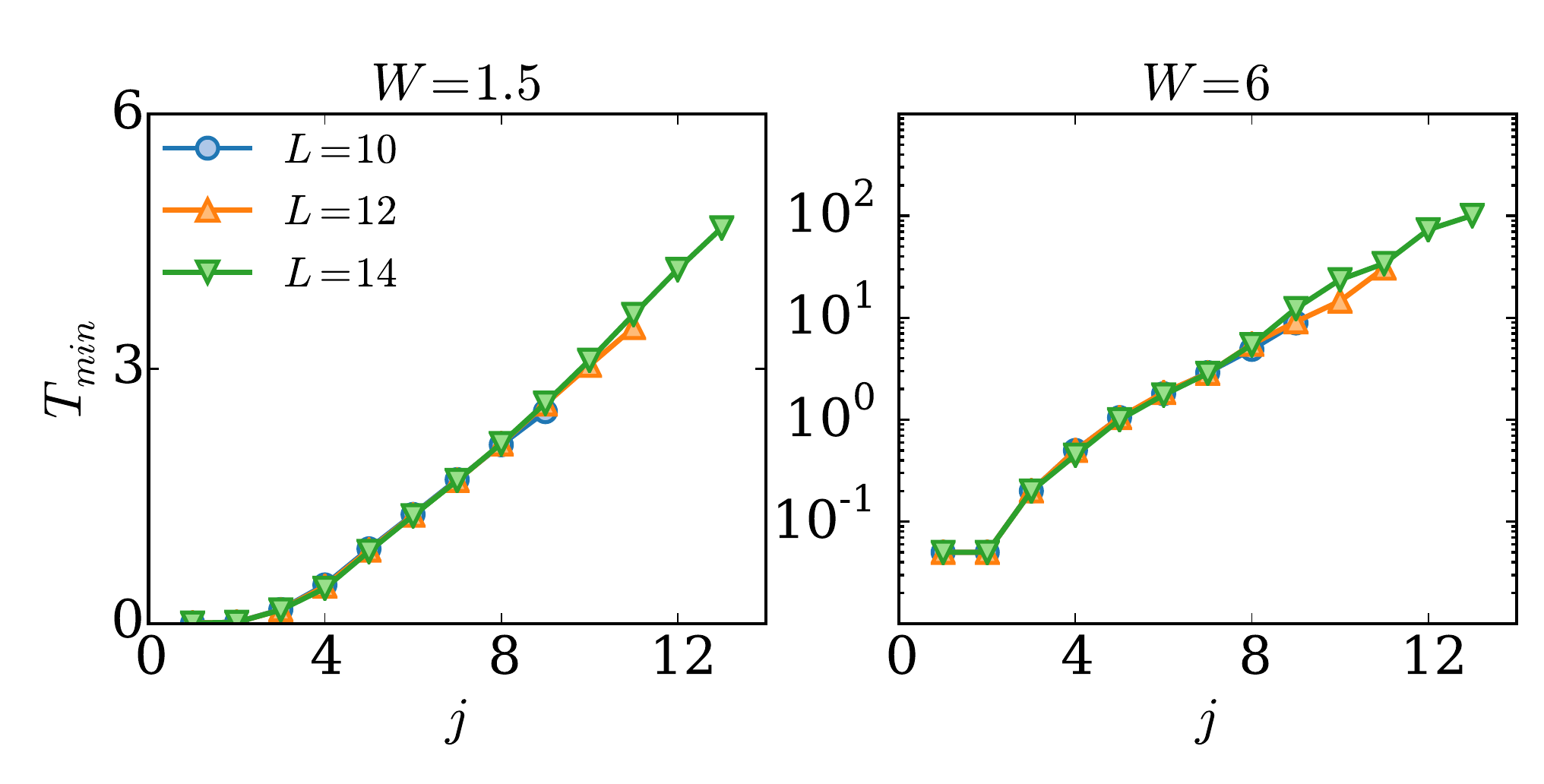}
\caption{$T_{\text{min}}$ for different system size and in two different phases.  For $W = 1.5$ extended phase, it grows 
algebraically. In
the localized phase $(W = 6)$ the time to entangled two separated region of the systems grows exponentially with their 
distance.}
\label{fig:Time1}
\end{figure}
\eq{ 
T_{\text{min}}(j) := \min \left \{ t  | \overline{\mathcal{I}_j}(t) \ge 10^{-5} \right \}
}
and we plot it as a function of $j$ in Fig.~\ref{fig:Time1}. 
In the extended phase (Fig.~\ref{fig:Time1}, left panel) $T_{\text{min}}$ grows algebraically with distance $j$, while in the MBL phase (Fig.~\ref{fig:Time1}, right panel) the time to entangle
two separated portions of the system grows exponentially with their distance after an intermediate regime.

{\it Conclusions}---In this work we studied the QMI in fermionic systems having a metal--insulator transition. First, 
we benchmarked our main conjectures on the scaling
of the QMI as a function of the distance of two sites in the AAH-model. Second, we studied it in a interacting model
having an MBL transition. The QMI decays exponentially with the distance in the localized phase and slower than exponential in the extended phase. 
This allowed us to define a correlation length $\xi$, which is finite in the localized phase and diverging in the extended phase. This correlation length recovers
the single particle localization length $\xi_{\text{loc}}$  if the system is composed of only a single fermion. Furthermore, we defined the quantity $\sigma$, which can be seen as the variance of an appropriate probability distribution defined using the quantum 
mutual information. In both models, this quantity saturates to a finite value in the localized phase and diverges with system size in the extended phase. 
Finally we studied the non-equilibrium properties of the MBL system by performing a global quench from a random product state and following the time evolution of the mutual information.
We showed that the spread of the QMI with time can be used as a dynamical indicator to distinguish an Anderson insulator phase from an MBL phase. In the Anderson phase it saturates with system size, while in the interacting case it 
grows logarithmically. With our study we propose the QMI between two sites as a possible quantity which in principle can be measured in experiments, to detect the MBL transition, and moreover to distinguish between an Anderson insulator phase and 
an interacting localized phase (MBL).
\begin{acknowledgments}
{\it Acknowledgments}---We thank J. Eisert, F. Heidrich-Meisner and V. Oganesyan for several illuminating discussions. 
We also express our gratitude to S. Vardhan for a critical reading of the manuscript. This work was 
partially supported by DFG Research Unit FOR 1807 through grants no. PO 1370/2-1 and by the ERC starting grant QUANTMATT NO. 679722.
\end{acknowledgments}

\bibliography{MutualBIB}


\clearpage

\subsection{Supplemental material to Quantum Mutual Information as a Probe for Many-Body Localization }
{\it Mutual information for two sites}---The quantum mutual information (QMI) between two sites $\{i,j\}$ in a fermionic system with a fixed number of particles is given by
\eq{ 
\mathcal{I}([i],[j]) := S([i]) + S([j]) - S([i] \cup [j])
}
where
\eq{\begin{split}
S([i]) = & - \langle n_i  \rangle \log \langle n_i \rangle  \\
&- \left (1- \langle n_i  \rangle \right)\log \left ( \langle 1 - n_i \rangle \right ),
\end{split}}
\eq{\begin{split}
S([i] \cup [j]) = & - \langle n_i n_j  \rangle \log \langle n_i n_j \rangle  \\
&-  \langle (1-n_i)(1-n_j)  \rangle\log  \langle (1 - n_i)(1-n_j) \rangle \\
& - \lambda_+ \log \lambda_+ -  \lambda_- \log \lambda_-,
\end{split}}
and
\eq{
\lambda_\pm = \frac{ -\langle (n_i + n_j)^2  \rangle \pm \sqrt{  \langle n_i - n_j  \rangle^2 + 4| \langle c_i^\dagger c_j \rangle |^2 } }{2},
}
where $\langle \cdot \rangle$ is the expectation value with an eigenstate of $\mathcal{H}$. The computation of the QMI requires only the knowledge of two point correlation functions (i.e. $\langle n_i  n_j  \rangle$) and the 
expectation values of the local densities ($\langle n_i  \rangle$).
In the case of one particle (N =1) the QMI reduces to
\begin{figure}[tb!]
\includegraphics[width=1.\columnwidth]{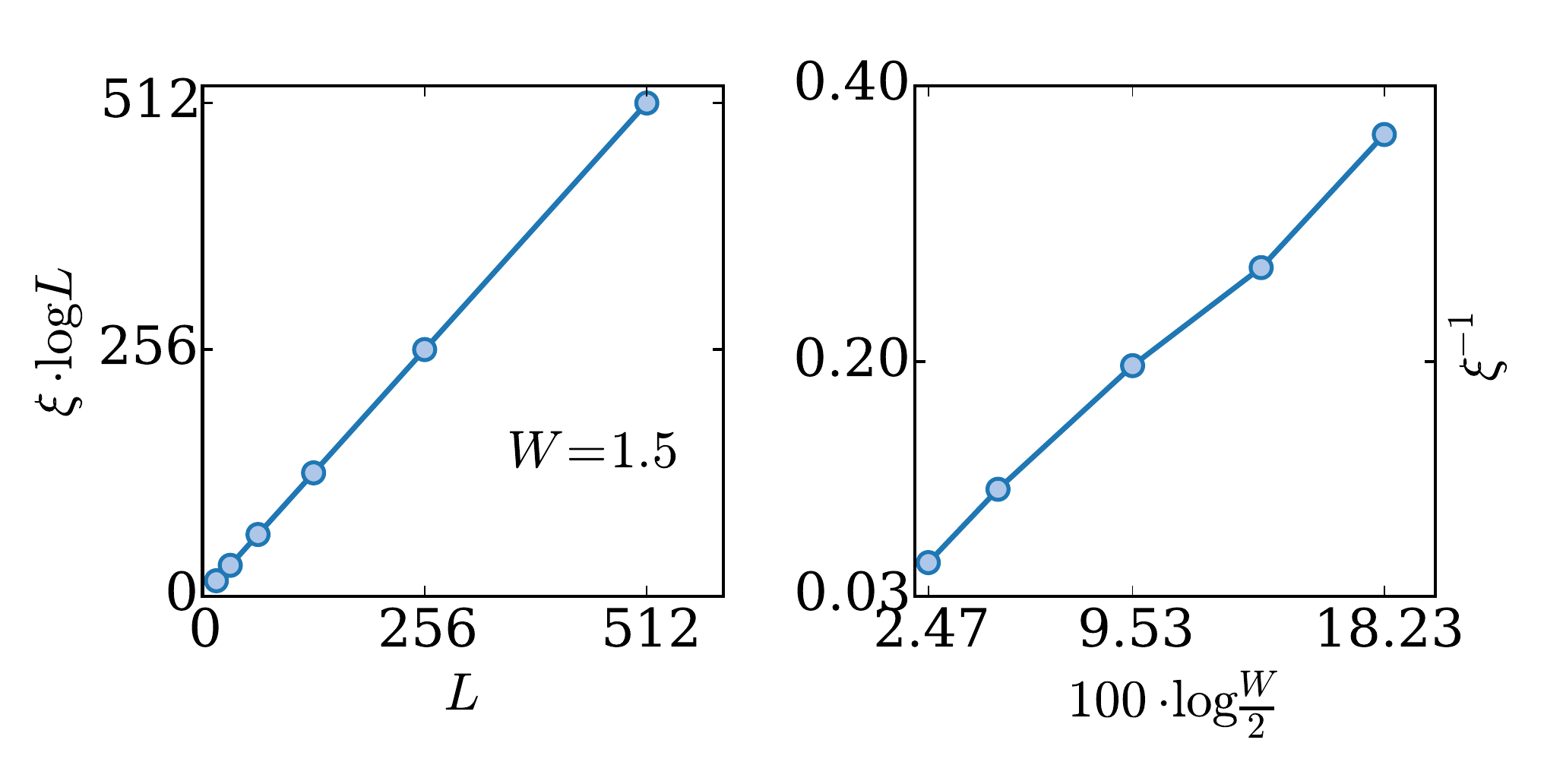}
\caption{(AAH -model) The left panel shows the localization length in the extended phase for the AAH model for 
different 
system sizes, $\xi \sim \frac{L}{\log L}$.  The right panel shows how $\xi$ approaches the transition point ($W_c =2$) 
as a function of $W$ in the localized phase. In the localized phase $\xi$ has been extrapolated choosing the system 
size 
$L$ in which $\xi$ saturates.}
\label{fig:xi_A}
\end{figure}
\begin{equation} 
\begin{split}
 \mathcal{I}_j  = & -|\psi_0|^2 \log |\psi_0|^2 - (1- |\psi_0|^2)\log (1-|\psi_0|^2) \\
 &  -|\psi_j|^2 \log |\psi_j|^2 - (1- |\psi_j|^2)\log (1-|\psi_j|^2) \\
 &  + (|\psi_0|^2+|\psi_j|^2 ) \log (|\psi_0|^2+|\psi_j|^2) \\
 & + (1-|\psi_0|^2-|\psi_j|^2 ) \log (1-|\psi_0|^2-|\psi_j|^2)
\end{split}
\end{equation}

\begin{figure}[tb!]
\includegraphics[width=1.\columnwidth]{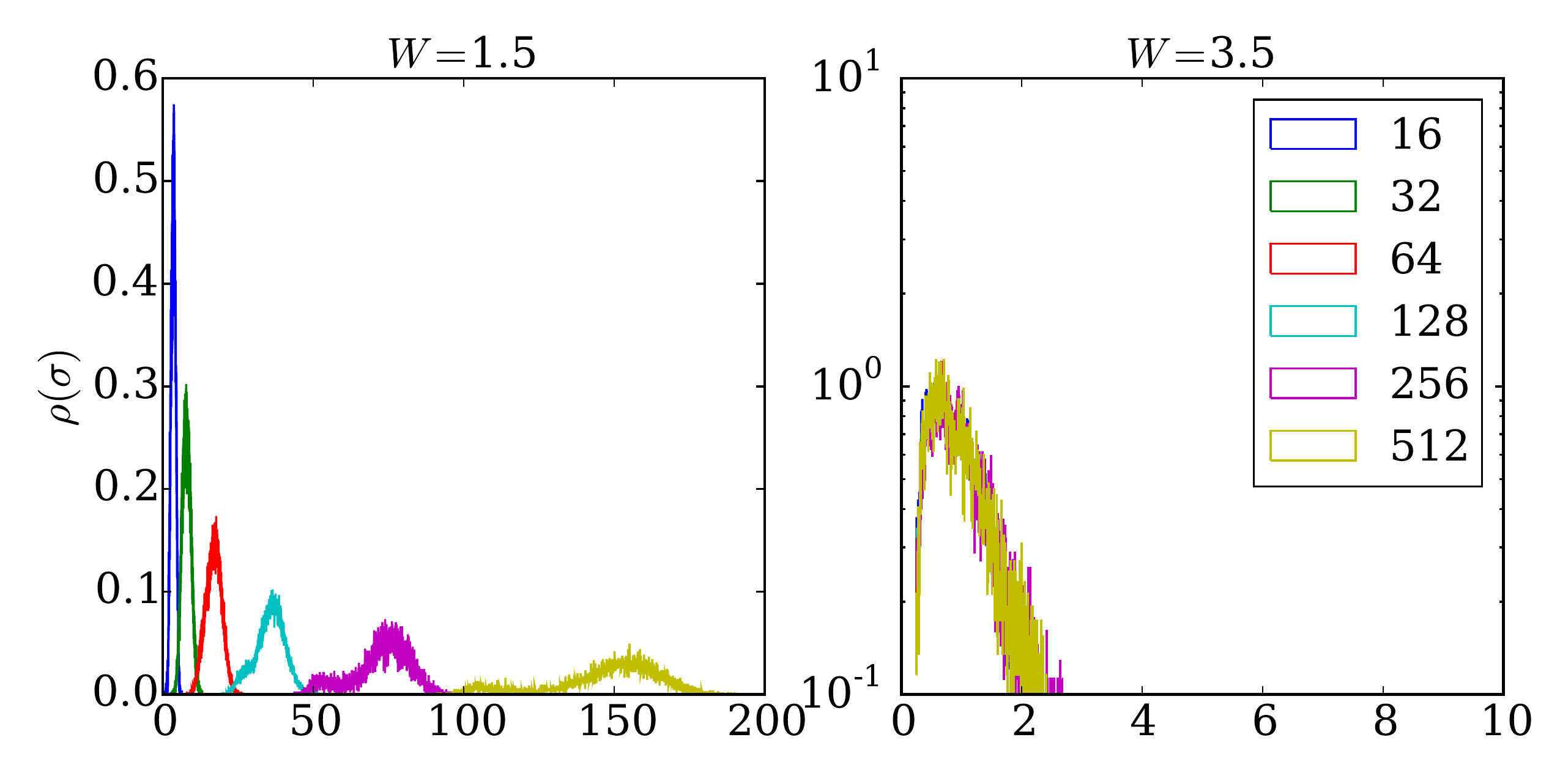}
\caption{(AAH -model) The probability distribution $\rho$ of $\sigma$ in the two different phase for different system 
sizes $L$. The first panel $(W =1.5)$ is in the extended phase and $\rho$ shifts to 
infinity with increasing $L$. The second panel $(W =3.5)$ is in the localized phase and $\rho$ does not scale with $L$. 
}
\label{fig:rho_A}
\end{figure}
{\it AAH-model}---In this section we report system size scaling for the QMI for the AAH model. The left panel of Fig.~\ref{fig:xi_A} shows how the correlation length $\xi$ grows with system size in the extended phase,  $\xi \sim \frac{L}{\log L}$. The logarithmic correction is due to the normalization of the single particle wave function in the extended phase, which decays as $\frac{1}{\sqrt{L}}$.
Moreover, the single particle localization length is known to diverge close to the critical point as $\xi_{\text{loc}} \sim \frac{1}{\log \frac{W}{2}}$. The right panel of Fig.~\ref{fig:xi_A} shows $\xi \sim \xi_{\text{loc}}$ close to the transition. 
It can be understood by the non existence of a single particle mobility edge in the AAH model, implying that the localization length of any particle diverges approaching $W_c$ as $\frac{1}{\log \frac{W}{2}}$, and thus the correlation length $\xi$ will be dominated by the 
divergence of $\xi_{\text{loc}}$.

Figure \ref{fig:rho_A} shows the full probability distribution ($\rho$) of $\sigma$ for the AAH-model in the two different phases. For $W=1.5$ in the extended phase, the probability shifts systematically with system size, indicating that all the states are extended.
In contrast, for $W=3.5$ in the localized phase $\rho$ does not shift, indicating that the system is fully localized.

\begin{figure}[tb!]
\includegraphics[width=1.\columnwidth]{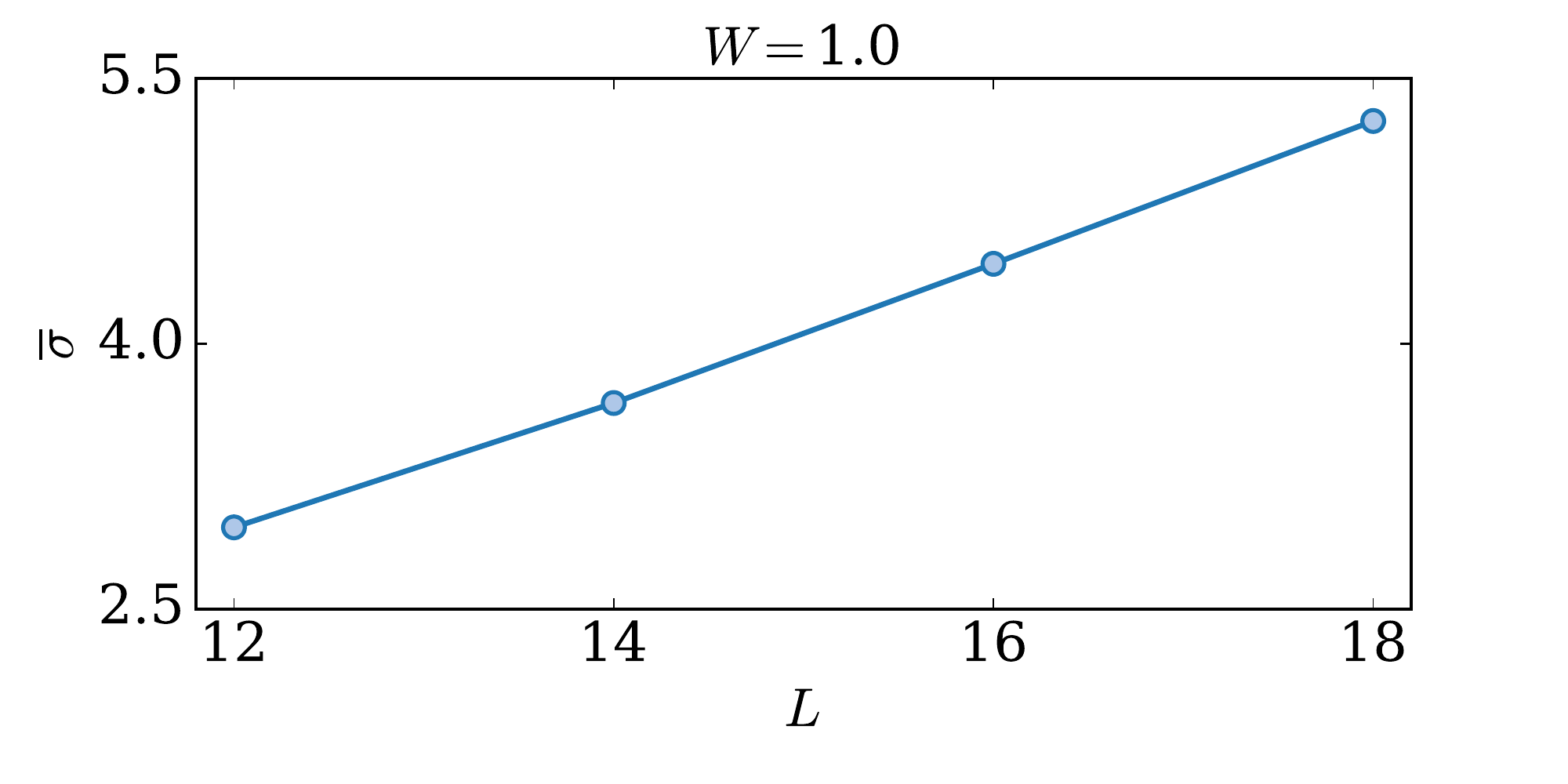}
\caption{(Spinless Hubbard chain) Scaling of $\overline{\sigma}$ for different system sizes in the extended phase ($W=1$).}
\label{fig:scaling_mbl}
\end{figure}

\begin{figure}[tb!]
\includegraphics[width=1.\columnwidth]{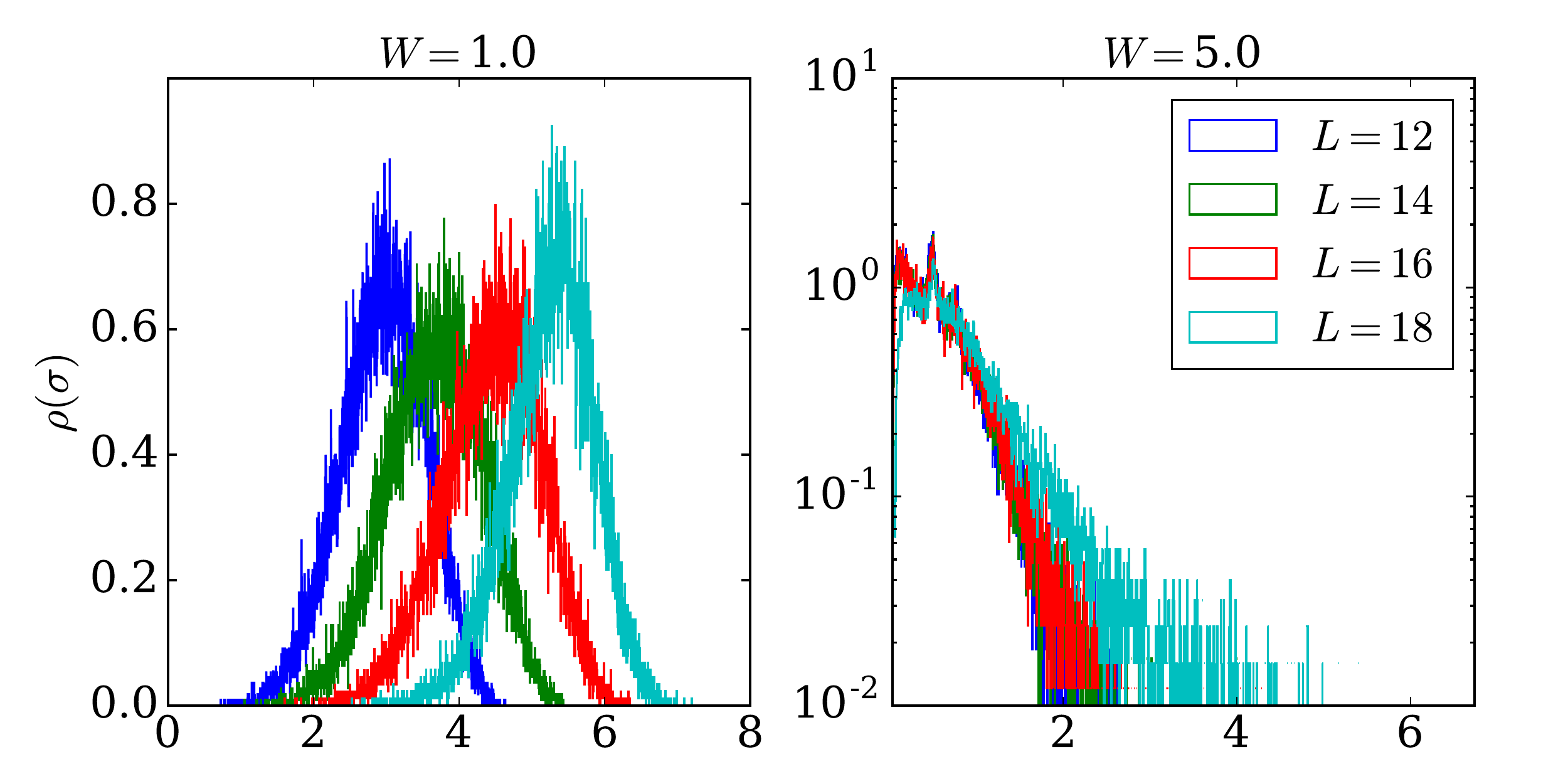}
\caption{(Spinless Hubbard chain) The probability distribution $\rho$ of $\sigma$ in the two different phases for different system sizes $L$. The first panel $(W =1.0)$ is in the extended phase and $\rho$ shifts to 
infinity with increasing $L$. The second panel $(W =5.0)$ is in the localized phase and it does not scale with $L$. }
\label{fig:rho_MBL}
\end{figure}

\begin{figure}
\includegraphics[width=1.\columnwidth]{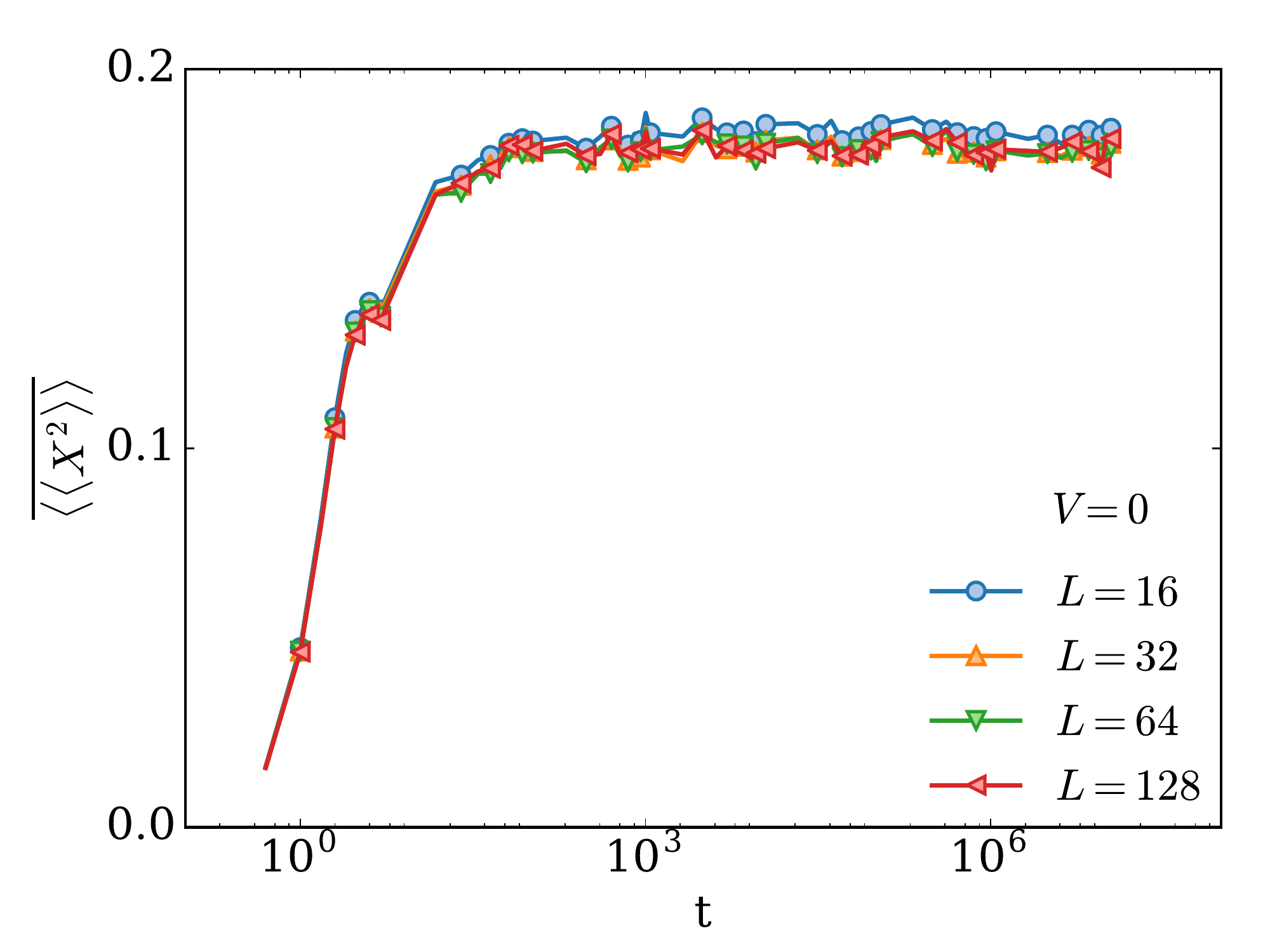}
\caption{(Spinless Hubbard chain) $\left \langle \langle X^2 \right \rangle \rangle$ as a function of time ($t$) for the Anderson model ($V=0$) for $W=6$.}
\label{fig:Anderson_time}
\end{figure}

{\it Spinless Hubbard chain}---Figure \ref{fig:scaling_mbl} shows the scaling of $\overline{\sigma}$ for different system sizes in the extended phase. $\overline{\sigma}$ scales linearly with $L$ indicating that $p_j \sim L^{-1}$, all sites
are correlated with each other uniformly. Figure \ref{fig:rho_MBL} shows the full probability distribution of $\sigma$ 
in the two different phases, in the extended phase ($W=1$) it shifts with system size, while in the MBL phase it is 
stable and has exponential tails.
Moreover, using a free fermion technique, we compute $\left \langle \langle X^2 \right \rangle \rangle$ for large system sizes for the noninteracting Anderson model ($V=0$), as shown in 
Fig.~\ref{fig:Anderson_time}. As expected $\left \langle \langle X^2 \right \rangle \rangle$ saturates with system size.
\end{document}